\documentclass[conference]{IEEEtran}
\IEEEoverridecommandlockouts
\usepackage{cite}
\usepackage{amsmath,amssymb,amsfonts}
\usepackage{graphicx}
\usepackage{textcomp}
\usepackage{xcolor}
\usepackage{amsmath,amsfonts}
\usepackage{graphicx}
\usepackage{textcomp}
\usepackage{booktabs}
\usepackage{makecell}

\usepackage{url}
\usepackage{multirow}
\usepackage{hyperref}

\usepackage{algorithm}
\usepackage{algpseudocode}
\usepackage{xcolor,soul}
\usepackage{makecell}

\usepackage{tikz}
\usetikzlibrary{shapes.geometric, arrows}

\usepackage{xcolor}

\newcommand{\scode}[1]{%
  \begingroup
  \definecolor{hlcolor}{gray}{0.93}\sethlcolor{hlcolor}%
  {\ttfamily \footnotesize #1}%
  \endgroup
}

\def\BibTeX{{\rm B\kern-.05em{\sc i\kern-.025em b}\kern-.08em
    T\kern-.1667em\lower.7ex\hbox{E}\kern-.125emX}}

\begin{document}

\title{Memory-Efficient Designs for Word-Wise Universal Fully Homomorphic Encryption}

\author{
\IEEEauthorblockN{Ardhi Wiratama Baskara Yudha\textsuperscript{\dag}, Erwin Eko Wahyudi\textsuperscript{\ddag}, Rian Adam Rajagede\textsuperscript{\ddag}, Qian Lou\textsuperscript{\ddag}, Yan Solihin\textsuperscript{\ddag}}
\IEEEauthorblockA{
\textsuperscript{\dag}\textit{Advanced Micro Devices, Inc., USA} \\
\textsuperscript{\ddag}\textit{University of Central Florida, USA} \\
Email: ardyudha@amd.com, \{wahyudierwin, rian, qian.lou, yan.solihin\}@ucf.edu
}
}

\maketitle

\makeatletter
\makeatother

\begin{abstract}

Fully Homomorphic Encryption (FHE) enables computation on encrypted data, preserving privacy throughout analysis. While its privacy is very strong, FHE is much slower to execute than the original computation. In particular, due to the recent success in accelerating its compute, the performance bottleneck shifts to the memory, especially considering that FHE magnifies the data size by orders of magnitude, resulting in a low arithmetic intensity. 

We propose BXT, an FHE optimization framework that mitigates the memory bottleneck through four techniques: (1) ciphertext compression, which regenerates ciphertext components from seeds during execution; (2) ciphertext serialization, which packs coefficients as bit arrays and unpacks them during L2-to-L1 transfer; (3) delayed seed generation, which defers PRNG-heavy offline work across aggregated operations; and (4) ciphertext digit pruning guided by fault-aware training tailored for Universal FHE. On CNN inference, the BXT-CSO50 configuration effectively achieves up to 3.8$\times$ speedup over the 100x GPU baseline with less than 1\% accuracy loss at 50\% comparison precision.

\end{abstract}


\section{Introduction}

\textit{Fully Homomorphic Encryption} (FHE)~\cite{Gentry2012FHEPolylog} is a cryptographic technique that permits computation directly on encrypted data. It enables collaboration among non-trusting parties. A client keeps its data private by sending encrypted data to the server, the server computes on the encrypted data while keeping its own computation secret, and the client receives the computation result in encrypted form, which can only be decrypted by the client itself. Hence, FHE allows private machine learning as a service (MLaaS), making FHE a key primitive for privacy-preserving cloud computing~\cite{AmazonScience, TensorFHE, reagen2021cheetah}.

FHE schemes differ in the native data and operations they support: word-wise schemes such as BGV~\cite{brakerski2014leveled} and CKKS~\cite{CKKS_scheme} support integers or floating-point numbers and arithmetic operations, while bit-wise schemes such as TFHE~\cite{TFHE_scheme} support logic operations. They are efficient for their native operations but inefficient for non-native ones; for example, word-wise schemes are inefficient for comparisons, while bit-wise schemes are inefficient for arithmetic. Recently, a BGV-based universal FHE (uFHE) scheme~\cite{iliashenko2021faster} enabled word-wise exact comparisons alongside arithmetic, making uFHE the state of the art for mixed operations with SIMD batching support, which TFHE lacks.

Regardless of the scheme, in general FHE incurs enormous {\em ciphertext expansion rate} (CER), i.e., the ciphertext is magnitudes larger than the original data. Hence, they typically show a low arithmetic intensity, i.e., few operations per byte accessed in memory (typically less than one~\cite{de2021does, 100x, mad23}). Many techniques have been proposed to accelerate FHE computation, but their success has made the memory bottleneck (hence arithmetic intensity) even worse in relative terms.

Recognizing the arithmetic intensity problem, an option to consider is to increase the GPU memory capacity or bandwidth. However, while this may accommodate FHE better, it is a costly approach. Recently, some have proposed packing multiple messages into a single ciphertext to improve slot utilization~\cite{100x, mad23}. None of these approaches changes the underlying inefficiency of the FHE ciphertext representations. In contrast, we propose mechanisms that directly reduce the ciphertext expansion and improve the arithmetic intensity. 

To this end, we propose \textit{BXT}, a novel multi-level optimization framework that targets the memory bottlenecks of FHE on GPUs. BXT introduces four techniques with complementary benefits:
\begin{itemize}
  \item \textbf{Ciphertext compression.} BXT-C uses pseudo-random number generator (PRNG)-based ciphertext compression to regenerate ciphertext components from compact seeds during execution. BXT regenerates ciphertext matrices dynamically in the GPU memory hierarchy, reducing CER and memory traffic.  
  \item \textbf{Ciphertext serialization.} BXT-CS additionally serializes ciphertexts by packing coefficients as bit arrays and unpacking them during L2-to-L1 cache transfer. This layout improves cache efficiency and reduces memory bandwidth demand without redesigning arithmetic units. 
  \item \textbf{Delayed seed generation.} BXT-CSO further applies delayed seed generation by deferring PRNG-heavy offline work and aggregating it across multiple operations. This strategy reduces intermediate matrix loads and stores, thereby reducing memory accesses.
  \item \textbf{Ciphertext digit pruning with fault-aware training.} For the uFHE scheme~\cite{iliashenko2021faster}, we discover an application-level opportunity. We create BXT-CSO75, BXT-CSO50, and BXT-CSO25 variants for pruning low-significance digits in comparisons, guided by fault-aware training, to reduce comparison precision while maintaining accuracy. This approach is tailored to digit-wise encrypted operations and provides tunable accuracy-performance trade-offs.

\end{itemize}

\textcolor{black}{We model and evaluate BXT using Accel-Sim~\cite{accel-sim}} and compare it against several state-of-the-art FHE systems on GPUs: 100x~\cite{100x}, TensorFHE~\cite{TensorFHE}, and GME~\cite{gme2023}. BXT increases arithmetic intensity by 1.5$\times$ and achieves up to 3.8$\times$ speedup over the 100x baseline~\cite{100x}, with less than 1\% accuracy loss at 50\% comparison precision. When integrated with GME~\cite{gme2023}, BXT speeds it up by 2.3$\times$.

\section{Related Work}

\textbf{GPU-based Acceleration of Operations.} The studies in~\cite{100x,TensorFHE,gme2023,HE-Booster,kim2024cheddarswiftfullyhomomorphic,Deng2024Trinity,wang2024chameleonefficientfhescheme} introduce solutions for accelerating FHE using GPUs (GPGPU).
100x~\cite{100x} is the first high-performance FHE implementation on GPGPU, which supports bootstrapping for CKKS but is constrained by inadequate hardware support for modulo calculations. TensorFHE~\cite{TensorFHE} enhances FHE arithmetic operations through algorithmic, Number Theoretic Transform (NTT), and data layout optimizations, and additionally leverages tensor cores to expedite NTT operations. GME~\cite{gme2023} presents a novel network-on-chip (NoC) that connects all scratchpad memory within the GPU, minimizing main memory accesses during NTT operations. HE-Booster~\cite{HE-Booster} refines these FHE operations by optimizing the GPU's NTT computation, building on~\cite{Ozgun2022EfficientNTT} with fine-grained synchronization at each NTT iteration. Recent works~\cite{Deng2024Trinity,wang2024chameleonefficientfhescheme} present effective schemes switching between TFHE and CKKS on GPUs. All these accelerators focus on arithmetic kernels, NTT optimization, data locality, or scheme switching. None of them reduces ciphertext size at runtime in memory, in contrast to BXT, which compresses ciphertexts and rebalances computation and memory traffic. Because BXT is orthogonal to these GPU accelerators, our techniques can be applied on top of existing GPU-based FHE implementations.

\noindent\textbf{ASIC/FPGA-based Acceleration of Operations.} Research such as~\cite{SHARP_ISCA23,ARK,BTS,F1,CraterLake,FxHENN} focuses on integrating an NTT unit optimized for radix-2 operations into FHE accelerators. CraterLake~\cite{CraterLake} stands out as the first FHE accelerator designed for high performance across a broad range of unbounded FHE programs, surpassing earlier designs that targeted a limited range of FHE computations~\cite{F1}. It features a uniprocessor setup with specialized units covering extensive vector spaces, employing static scheduling to leverage the regular patterns in FHE computations. SHARP~\cite{SHARP_ISCA23} aims to cut FHE operation latency by restricting the prime modulus to just 36 bits, which reduces the memory bandwidth demands and consequently boosts performance.
While ASIC/FPGA platforms offer significant acceleration and power-efficiency advantages over GPUs, they typically require longer development time and more effort to adapt to new schemes, algorithms, and implementations. In this sense, GPUs provide greater flexibility for rapid deployment and iteration.

\noindent\textbf{Ciphertext Compression.} 
Ciphertext compression utilizing PRNG was proposed for reducing network communication costs~\cite{rasoul2024compressingfhe} and for offline key-switching matrix generation (CraterLake~\cite{CraterLake}, ARK~\cite{ARK}). We adopt this approach as one of the four techniques of BXT, but with a major difference in that BXT utilizes this for computation, i.e., ciphertext compression is performed at runtime for computation, and components are regenerated on-the-fly in the memory hierarchy from seeds to directly reduce the memory bandwidth consumption on GPUs.

\section{Background}
\subsection{Universal FHE and BGV Scheme}

Traditional arithmetic FHE schemes such as BGV~\cite{brakerski2014leveled} primarily support arithmetic operations. Recent enhancements~\cite{yudha2024boostcom, iliashenko2021faster} enable BGV to perform efficient word-wise comparisons while preserving its arithmetic capabilities, establishing it as a universal FHE (uFHE) solution. Alternative approaches such as TFHE-BGV~\cite{Chimera} and TFHE-CKKS~\cite{lu2021pegasus} require scheme-switching with over 70$\times$ latency overhead~\cite{iliashenko2021faster}, making the enhanced BGV the state-of-the-art uFHE.

BGV uses lattice-based encryption founded on the Ring Learning with Errors (RLWE) problem~\cite{brakerski2014leveled}. Key parameters include: $N$ (polynomial degree, typically $2^{16}$), $Q$ (ciphertext modulus as a product of primes $q_i$), $p$ (plaintext modulus), and $L$ (multiplicative depth). Polynomials operate in the ring $R_Q = \mathbb{Z}_Q[x]/(\Phi_m(x))$, where $\Phi_m(x)$ is the $m$-th cyclotomic polynomial.

\textbf{Encryption.} The secret key $S \in R_Q$ is a small polynomial with coefficients in $\{-1, 0, 1\}$. To encrypt a plaintext $\mu \in R_p$, the scheme samples a random polynomial $A$ from $R_Q$ using a seed, and computes $B = -A \cdot S + p \cdot E + \mu$, where $E$ is sampled from a Gaussian distribution. The ciphertext is $C = (A, B)$. Crucially, $A$ can be regenerated from its seed, enabling compression. Decryption computes $\mu = ((B + A \cdot S) \bmod Q) \bmod p$.

\textbf{Polynomial Representation \& NTT.} Ciphertexts consist of polynomial pairs with a degree of $N$ and coefficients of size $\log(Q)$. To manage large $Q$, BGV uses the Residue Number System (RNS) via the Chinese Remainder Theorem, dividing polynomials into $L+1$ residue polynomials with coefficients under moduli $q_i$, forming a matrix of size $(L+1) \times m$. The Number Theoretic Transform (NTT) enables efficient polynomial multiplication in $O(N \log N)$ time~\cite{Cooley1965FFT}. FHE libraries~\cite{HElib, OpenFHE} use NTT with RNS, known as the DoubleCRT format~\cite{HElib}, to represent polynomial coefficients as word-sized integers.

BGV supports five key homomorphic operations: PADD (plaintext addition), PMULT (plaintext multiplication), HADD (ciphertext addition), HMULT (ciphertext multiplication), and HROTATE (ciphertext rotation). HADD and PADD use element-wise addition, PMULT uses Hadamard multiplication, while HMULT and HROTATE require NTT, basis conversion, and key switching~\cite{Gentry2012FHEPolylog}, making them computationally intensive.

\subsection{Integer-wise Comparison}
\label{sec:he-comparison}
To compare two encrypted integers $\mathbf{x}$ and $\mathbf{y}$, each integer is first decomposed into $l$ digits over $\mathbb{F}_{p}$. Comparison is then performed digit-wise using polynomial interpolation~\cite{iliashenko2021faster} to evaluate the equality ($EQ$) and less-than ($LT$) functions on each pair of corresponding digits independently. Specifically, ${EQ(x, y)}$ equals 1 if \(x = y\), and 0 otherwise, while ${LT(x, y)}$ equals 1 if \(x < y\), and 0 otherwise. In the final step, the digit-level results are aggregated lexicographically:
$EQ(\mathbf{x},\mathbf{y}) = \prod_{i=0}^{l-1} EQ(x_i,y_i)$ and
$LT(\mathbf{x},\mathbf{y}) = \sum_{i=0}^{l-1} LT(x_i,y_i) \prod_{j=i+1}^{l-1} EQ(x_j,y_j)$.
This digit-wise comparison method substantially increases the number of ciphertexts and the overall memory footprint compared to arithmetic-only FHE.
\section{Motivation}

\begin{figure}[htbp]
\centering
\includegraphics[width=0.95\linewidth]{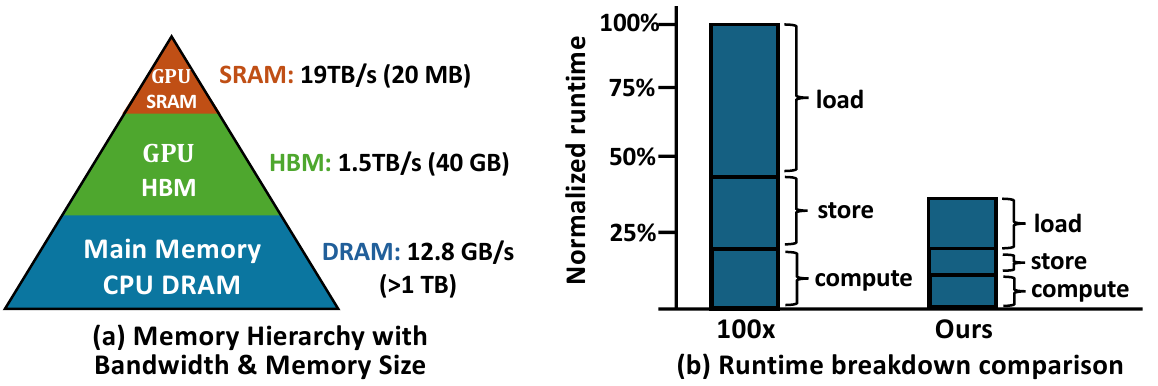}
\caption{(a) GPU memory hierarchy and (b) access inefficiencies in the GPU-based FHE accelerator (100×) highlight the need for memory-efficient uFHE designs.}
\label{fig:motivation}
\vspace{-1ex}
\end{figure}

The GPU memory hierarchy in Figure~\ref{fig:motivation}(a) consists of multiple memory types with varying sizes and speeds. NVIDIA's A100 GPU features 40--80 GB of HBM (1.5--2.0 TB/s bandwidth) and 192 KB of on-chip SRAM per SM (aggregate \(\sim\)19 TB/s). While SRAM is significantly faster than HBM, it is orders of magnitude smaller. As compute performance has outpaced memory speed improvements, memory has become a major bottleneck, making efficient use of on-chip SRAM critical.

As shown in Figure~\ref{fig:motivation}(b), when the prior GPU-based FHE accelerator (100×) is adapted for uFHE comparison operations, memory operations account for over 80\% of runtime. In uFHE, each digit is encrypted into a separate ciphertext, significantly increasing CER. This overhead is pronounced in DNN workloads with hundreds of non-linear comparisons (e.g., ReLU layers). The high CER leads to low ArI~\cite{de2021does, 100x,mad23}, with fewer than one arithmetic operation per byte, causing frequent memory accesses and limited data reuse. We address this through four techniques: (1) compressing ciphertexts, (2) serializing coefficients to remove redundant bits, (3) delayed seed generation, and (4) pruning less significant digits.

FHE inherently expands data size, requiring large parameters for bootstrapping and security. Encrypted data can grow by hundreds of times, depending on parameters and batching utilization. Even with full batching, ciphertexts often exceed cache capacity. Limited cache space hampers locality, leading to frequent memory accesses. Memory access time dominates over cache and ALU execution combined, making memory load latency reduction essential.

\section{BXT Design}


To effectively address the issue of high CER and enhance computation efficiency, we introduce BXT, a novel multi-level optimization strategy spanning four levels: algorithm (ciphertext compression via seed regeneration), data layout (bit-array storage), operation (delayed seed generation), and application (digit reduction in comparisons).

\subsection{Ciphertext Compression and Serialization}

\begin{figure}[htbp]
\vspace{-1ex}
\centering
\includegraphics[width=0.95\linewidth]{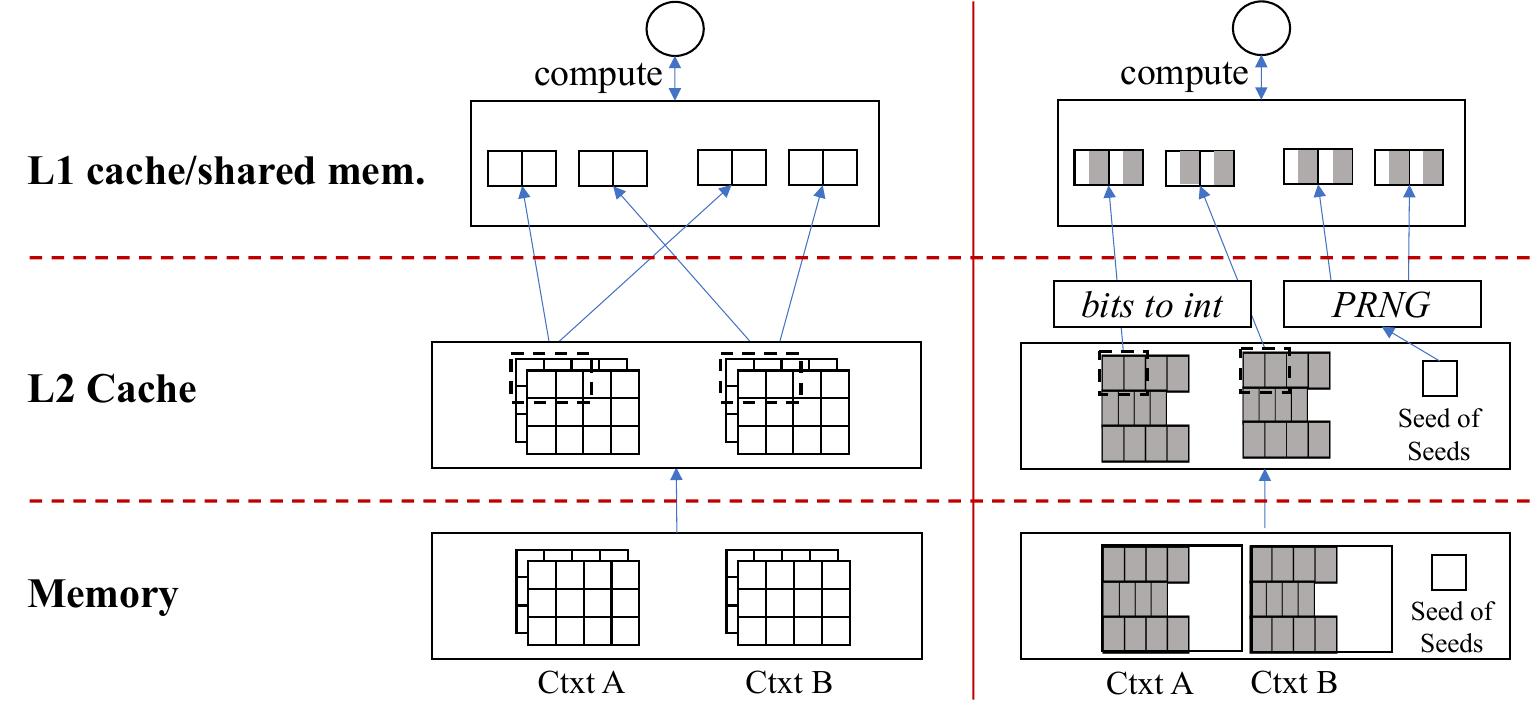}
\vspace{-1ex}
\caption{Data flow for processing ciphertext. (Left) Data is fetched from memory through the L2 cache to the L1 cache. (Right) The ciphertext is compressed and serialized into a bit array, which is then converted to integers. A PRNG generates the read-only ciphertext portion from a seed.}
\label{fig:dataflow-ciphertext-compression}
\vspace{-1ex}
\end{figure}

A ciphertext comprises two matrices stored in memory. Operating on two ciphertexts requires fetching all four matrices from memory to the L2 cache and then transferring portions of them to the L1 cache for processing (Figure~\ref{fig:dataflow-ciphertext-compression}, left). With matrix sizes being large (e.g., ranging from $2^{13}$ to $2^{16}$ for security strength), fetching them demands significant memory bandwidth and reduces arithmetic intensity.

To address bandwidth issues, we leverage the fact that one ciphertext matrix is randomly generated. We store only the computed matrix and a seed, regenerating the other matrix on-the-fly using a hardware-supported pseudo-random number generator (PRNG). Similar techniques were utilized for reducing the message size of ciphertext communication~\cite{rasoul2024compressingfhe} or for offline key-switching matrix generation (CraterLake~\cite{CraterLake}, ARK~\cite{ARK}). We adopt this approach for storing the ciphertext in memory and fetching it from memory, resulting in halving the storage requirements and reducing the memory bandwidth. 

\textbf{Security Considerations.}
We assume the same threat model as the underlying BGV-based uFHE scheme: the adversary may observe public parameters, public keys, and ciphertexts stored in memory, but does not know the secret key. As with~\cite{rasoul2024compressingfhe}, we assume a PRNG producing uniformly distributed random numbers from a seed that is independent from the secret key. In BXT, the seed is used only to regenerate the random ciphertext component, rather than storing it explicitly, in order to save space. If an adversary learns the seed, they can reconstruct this regenerated random component, but this alone does not reveal the plaintext or the secret key under the standard RLWE-based security assumption. Thus, BXT does not alter the underlying cryptographic security model; it only changes how part of the ciphertext is stored and reconstructed during execution.

\textbf{Ciphertext serialization.}
We further optimize memory through data layout reorganization. Matrix elements use moduli typically up to 36 bits~\cite{SHARP_ISCA23}, smaller than 64 bits. By storing elements as bit arrays instead of full integers, we eliminate unused bits, reduce the matrix sizes, and hence the memory bandwidth. 

One key design question is when to unpack the bits. One option is to unpack them at the load/store unit of the processor. This saves the storage in the entire cache hierarchy, but complicates the pipeline design due to the unpredictable load/store unit timing. 
The opposite option is to unpack them as soon as they enter the chip, before storing them in the L2 cache. This reduces the opportunity to also save space in the L2 cache. Thus, instead, we unpack during the L2-to-L1 transfer, avoiding massive compute-unit redesign while improving cache efficiency. Figure~\ref{fig:dataflow-ciphertext-compression} (right) illustrates the optimized data flow: compressed data is fetched to L2 cache, converted from bit arrays to integers, and stored in L1 cache for GPU computation. For read-only ciphertexts, one matrix is stored while the other is generated on-the-fly via PRNG (preferably hardware-backed for lower latency). We then introduce new instructions for loading and storing ciphertexts at matrix-fragment granularity, similar to Tensor Cores~\cite{Markidis2018TensorCore}, enabling efficient bit-to-integer conversion.

\subsubsection{Load/Store Interface for Read-Only Ciphertext}
For read-only ciphertexts, we optimize storage by generating one matrix dynamically from a seed while storing only the other in memory. We introduce new load/store instructions inspired by Tensor Core matrix fragments~\cite{Markidis2018TensorCore} that enable on-the-fly generation at warp-level granularity, allowing 32 elements to be fetched and generated simultaneously, as outlined in Table \ref{tab:PRNG_APIs}.

\begin{table}[htbp]
\vspace{-1.5ex}
\caption{Proposed Pseudo-Random Number Generator Instructions.}
\label{tab:PRNG_APIs}
\centering
\begin{tabular}{|>{\raggedright\arraybackslash}p{8cm}|} 
\hline
\textbf{Variable declaration and interfaces for load/store PRNG-generated matrix}\\ \hline \hline
\scode{wmma::fragment<unused, num\_of\_element, unused, unused, unused, unused> A\_frag} \\ \hline
\scode{wmma::load\_limb\_fragment\_prng(A\_frag, seed+index, modulus);}
\\ \hline 
\scode{wmma::store\_limb\_fragment\_prng(mem\_destination, A\_frag, modulus);}
\\ \hline
\end{tabular}
\vspace{-1ex}
\end{table}

The \scode{num\_of\_element} parameter in the variable declaration specifies the number of elements in the fragment limb to be generated, allowing operations to be executed with warp-level granularity. Given that the current FHE implementation fetches one element of a ciphertext matrix per thread, setting \scode{num\_of\_element} to 32 allows for fetching and generating 32 elements simultaneously, thereby enhancing efficiency.

\subsubsection{Load/Store Interface for Data Layout Modification}
The matrix stored in memory is represented as an array of bits, requiring conversion to recover integers for computation. To facilitate this conversion, we introduce specific variable declarations and load/store instructions, as detailed in Table \ref{tab:CC_APIs}.

\begin{table}[htbp]
\vspace{-1.5ex}
\caption{Proposed Data Layout Converter Instructions.}
\label{tab:CC_APIs}
\centering
\begin{tabular}{|>{\raggedright\arraybackslash}p{8cm}|} 
\hline
\textbf{Variable declaration and interface for load/store compressed ciphertext}\\ \hline \hline
\scode{wmma::fragment<unused, num\_of\_element, unused, unused, unused, unused> A\_frag} \\ \hline
\scode{wmma::load\_limb\_fragment(A\_frag, mem\_source, log2(modulus));}
\\ \hline
\scode{wmma::store\_limb\_fragment(mem\_destination, A\_frag, log2(modulus));}
\\ \hline
\end{tabular}
\vspace{-3ex}
\end{table}

As depicted in the table, we begin by declaring a variable to hold the fetched matrix elements, specifying the exact number of elements to be retrieved. This initial declaration sets up the framework for efficient data handling and ensures that the matrix elements are fetched in the required format. Subsequently, the load/store instructions are employed to retrieve the array of bits and transform the layout into 64-bit integers. This layout transformation is facilitated by dedicated hardware positioned between the L1 and L2 caches. Notably, the number of bits to be converted into a single 64-bit integer is specified as the final argument in the new load/store instructions.

\subsection{Ciphertext Operation Optimization}

For operations such as PMULT, PADD, and HADD, performance is primarily limited by matrix loads and stores rather than computation. For example, in HADD, 67\% of the execution time is spent on matrix loads and 29\% on matrix stores, while only about 4\% is used for computation, kernel launch, and modulus data fetching. This breakdown shows that reducing matrix load and store overhead is critical to lowering latency and improving overall performance.

\begin{figure}[htbp]
\centering
\vspace{-1ex}
\includegraphics[width=0.95\linewidth]{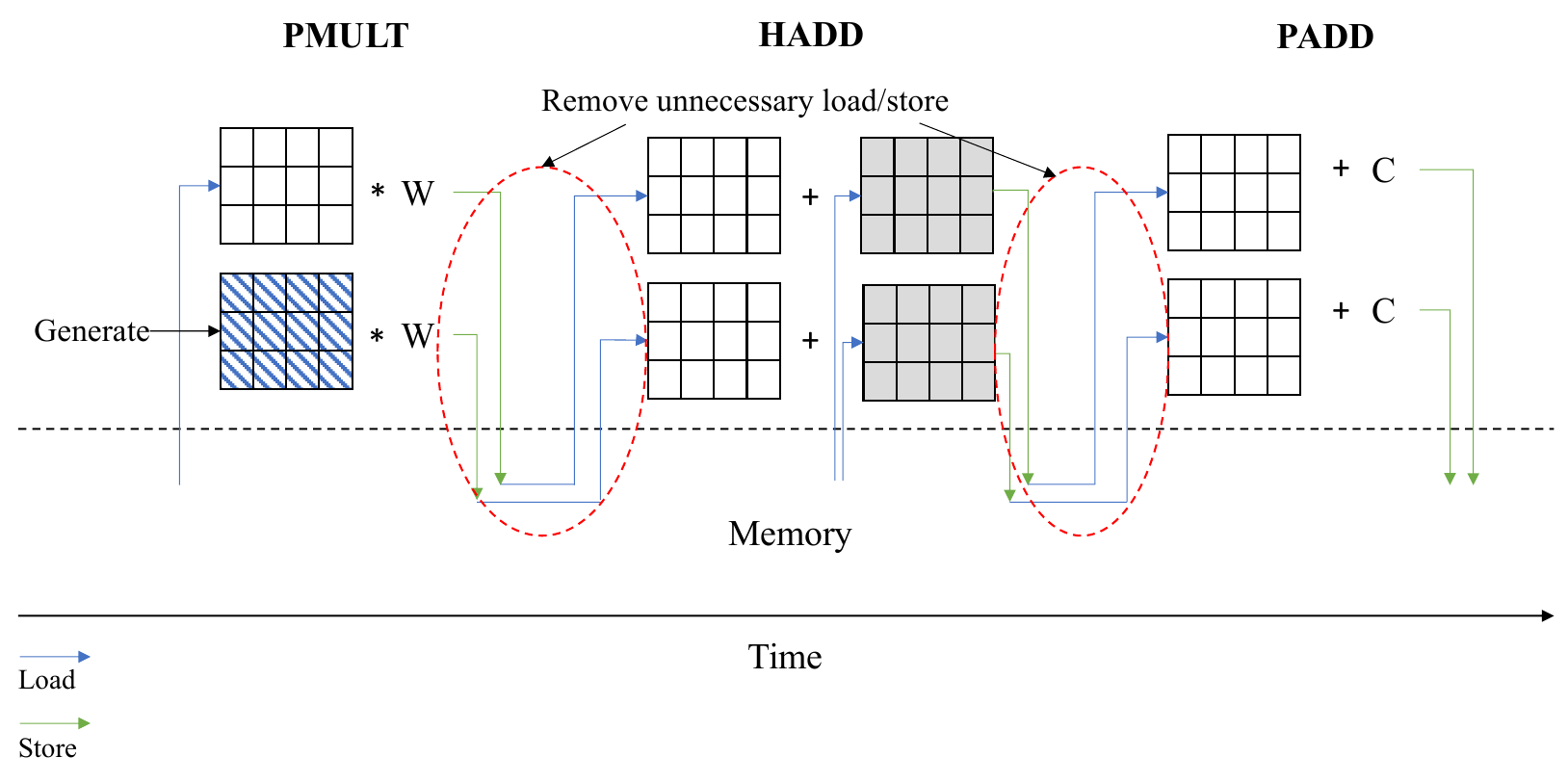}
\vspace{-1ex}
\caption{Timeline diagram depicting the matrix loads and stores for the sequential operations of PMULT, HADD, and PADD. Combining these operations will reduce the number of loads/stores for the matrix, as highlighted by the red oval lines.}
\label{fig:unfused-ops}
\vspace{-1ex}
\end{figure}

We propose aggregating common operation sequences such as PMULT, HADD, and PADD to reduce data movement overhead. This optimization is motivated by their frequent use in convolutional and fully connected layers in neural networks. For example, to compute \(x \cdot w + b\), where \(x\) is ciphertext and \(w\) and \(b\) are plaintext, PMULT and HADD are first used to compute \(x \cdot w\), followed by PADD to add \(b\). As shown in Figure~\ref{fig:unfused-ops}, aggregation removes the intermediate memory transfers highlighted by the red dashed ovals, reducing matrix loads and stores and thus improving efficiency and latency.

\begin{figure}[htbp]
\vspace{-1ex}
\centering
\includegraphics[width=0.95\linewidth]{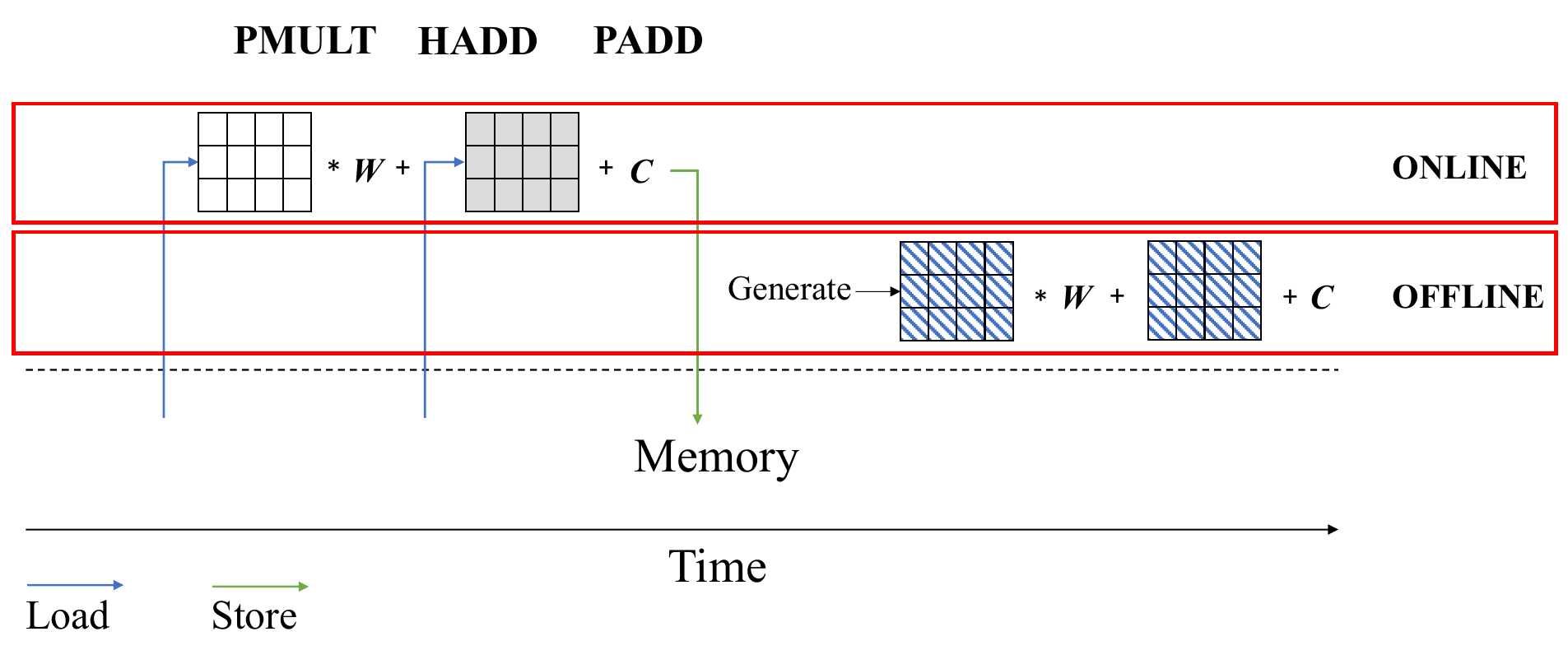}
\vspace{-0.5ex}
\caption{Timeline diagram illustrating the online and offline stages of the PMULT, HADD, and PADD operations, with the offline parts delayed to conserve matrix loads.}
\label{fig:fused-ops}
\vspace{-3ex}
\end{figure}

We also propose another operation-level optimization, which we term delayed seed generation. As shown in Figure \ref{fig:fused-ops}, the stages of the PMULT, HADD, and PADD operations can be divided into online and offline stages. The online stage involves loading matrices from memory, while the offline stage bypasses matrix loads by generating data through a PRNG engine. Additionally, the offline parts of the operations can be postponed until needed, allowing them to be combined with the online parts in operations that require both matrices, such as HMULT or HROTATE operations. Delaying the offline stage in this way helps conserve matrix loads, thus reducing the execution times of the operations.
\textcolor{black}{After each operation, the output ciphertext is stored in an uncompressed form because the seed representation cannot be directly propagated through these operations.}

\subsection{Ciphertext Pruning}
\label{sec:red-prec}

Our last optimization is specific to a BGV-based universal FHE (uFHE) scheme~\cite{iliashenko2021faster} that enables fast word-wise exact comparisons by operating digit-wise. Without uFHE, comparison operations are very slow compared to arithmetic operations, and therefore various scheme switching techniques have been proposed. uFHE enables fast comparisons by operating on encrypted digits, but it requires each digit to be stored in a separate ciphertext.

\begin{figure}[htbp]
\centering
\includegraphics[width=0.95\linewidth]{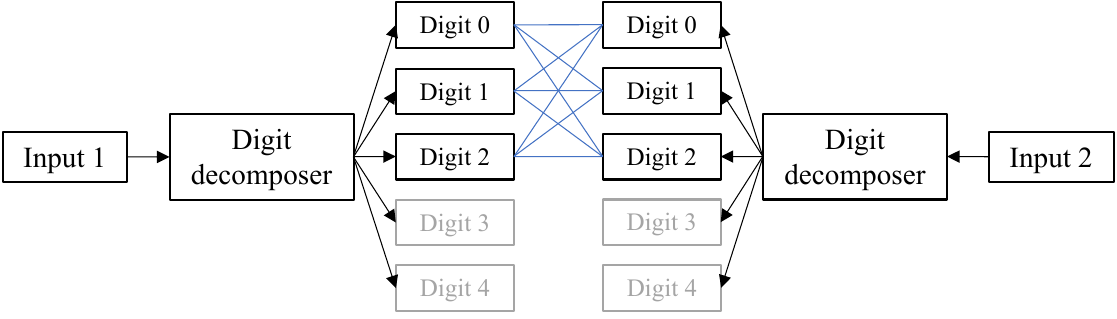}
\caption{Comparing two numbers begins with decomposing each number into digits. The digits colored black show the digits used in comparison, while the digits in gray are unused.}
\label{fig:reduced-precision-comparison}
\vspace{-1ex}
\end{figure}

As described in Section~\ref{sec:he-comparison}, uFHE performs comparisons in a digit-wise manner and then combines the digit-level results to produce the final comparison result. To speed up comparisons, we propose reducing the number of digits used. As shown in Figure~\ref{fig:reduced-precision-comparison}, the digit decomposer outputs five digits, yet only three are used in the comparison. The number of digits to retain is determined beforehand by the decomposer. By intentionally discarding low-significance digits, we reduce the effective comparison precision: fewer ciphertexts are stored and compared, resulting in a lower computational load and faster processing. Pruning and precision reduction are widely used in machine learning~\cite{michaud2023precisionML}, but applying reduced-precision pruning directly at the ciphertext-digit level for encrypted comparisons is, to our knowledge, novel.

However, it is important to acknowledge that disregarding certain digits can lead to errors in comparisons. In machine learning workloads, minor errors are typically acceptable if the resulting drop in inference accuracy is minimal and remains within acceptable limits. Identifying the optimal level of precision reduction that still yields reliable outcomes is essential. 

Furthermore, we mitigate the decreases in inference accuracy by training the network differently. In particular, we can regard ciphertext pruning as logically equivalent to the network operating in a faulty environment~\cite{zahid2020fat}. After all, disregarding several least significant digits may result in an incorrect comparison outcome, similar to when those same digits are affected by faults. Therefore, we can utilize \textit{fault-aware training} to add robustness to the network. This approach involves training the model with a custom ReLU function designed to produce the same erroneous outputs as those which may have been caused by digit reduction. This method allows the model to anticipate and correct these specific types of errors during its operation. To perform this training, we set a parameter $x$ that limits the faults to $(100-x)$\% of the least significant digits. For example, $x=75$ means only 25\% of the least significant digits are injected with faults.

We note that, as shown in many prior studies, there may be a trade-off between robustness and accuracy, that is, the more robust a network is against faults, the lower its accuracy. Hence, we need to find this sweet spot between accuracy preservation and robustness against ciphertext pruning. 

\section{Methodology}

\subsection{Evaluation Environment}

We extend BXT \textcolor{black}{starting with 100x~\cite{100x} as a base}. Then, we model it on Accel-sim~\cite{accel-sim} configured with Nvidia RTX A6000 architecture (Ampere architecture, Table~\ref{tab:gpu_config}). Each streaming multiprocessor (SM) has a 40-cycle PRNG~\cite{yuan2021analyzing} and a 2-cycle bit-to-int converter. 

\begin{table}[htbp]
\vspace{-1.5ex}
  \caption{GPU Configuration}
  \label{tab:gpu_config}
\centering
  {
  \begin{tabular}{|p{2.5cm}|p{5cm}|}
    \hline
     {\bf Aspect} & {\bf Configuration} \\
     \hline \hline 
     SM Configuration & 84 SMs, 1410 MHz \\
     \hline
     Register File &  256 KB/SM, 20.5 MB in total\\
     \hline
     Process Size & 8 nm \\
     \hline
     L1D and Shared Mem. & 128 KB \\
     \hline
     L2 cache & 2 banks per memory partition, each L2 cache bank is 128 KB, 6 MB in total. \\
     \hline
     DRAM & 48 GB at 1219 MHz, 24 partitions, 936.2 GB/s\\
    \hline 
  \end{tabular}}
\end{table}

We employ FHE parameters from \cite{gme2023} that are conducive to bootstrapping and provide 128-bit security ($\lambda=128$): the polynomial degree $N=2^{16}$, ciphertext modulus bit-length $\log(Q)=1728$, with $L=23$ levels for computation and $L_{boot}=17$ for bootstrapping, and a decomposition number $dnum=3$.

\subsection{Workloads Evaluated}

We evaluate five homomorphic operations (PADD, PMULT, HADD, HMULT, HROTATE) using BGV kernels~\cite{100x}, and two ML workloads on encrypted MNIST data~\cite{lecun1998mnist}:

\textbf{Logistic Regression} classifies digits 3 vs.\ 8. Each $28\times28$ image is encrypted into 784 ciphertexts (one per pixel). Inference computes $S = \sum w_i x_i$ with plaintext weights, outputting 1 if $S \geq 0$, and 0 otherwise.

\textbf{CNN} uses five layers from CryptoDL~\cite{cryptodl}: Conv ($5\times5$ kernels, stride 2), ReLU ($ReLU(x) = x \cdot (1 - LT(x, 0))$), FC (100 nodes), ReLU, and Output (10 nodes). Unlike prior work using polynomial approximations~\cite{HE-Booster}, we employ exact ReLU via encrypted comparison.

\subsection{Schemes Evaluated}


Table~\ref{tab:bxt-schemes} lists our BXT variants arranged based on increasing number of optimizations (BXT-C, BXT-CS, and BXT-CSO) and whether ciphertext pruning at various levels is applied (BXT-CSO75, BXT-CSO50, and BXT-CSO25).
\begin{table}[htbp]
\vspace{-1.5ex}
  \caption{BXT Variants}
  \label{tab:bxt-schemes}
\centering
  \begin{tabular}{|l|p{6cm}|} \hline
     {\bf Scheme} & {\bf Brief Description} \\ \hline \hline 
     BXT-C & our scheme with PRNG optimization \\ \hline 
     BXT-CS & BXT-C plus bit-serialization layout optimization \\ \hline 
     BXT-CSO & BXT-CS plus operation optimization (delayed seed generation) \\ \hline 
     BXT-CSO75 & BXT-CSO with (100-75)\% ciphertext digits pruned  \\ \hline 
     BXT-CSO50 & BXT-CSO with (100-50)\% ciphertext digits pruned  \\ \hline 
     BXT-CSO25 & BXT-CSO with (100-25)\% ciphertext digits pruned  \\ \hline
  \end{tabular}
\end{table}

In addition to evaluating BXT variants' performance, we will put BXT's performance in the context of several state-of-the-art FHE systems on GPUs: 100x~\cite{100x}, TensorFHE~\cite{TensorFHE}, and GME~\cite{gme2023} (Section~\ref{sec:comparo}). We note that these systems do not perform ciphertext compression; therefore, BXT is orthogonal to them and can be integrated with them. To demonstrate this, we combine BXT with GME and show that their combined speedups greatly outweigh each one on its own.

Finally, fault-aware training (Section~\ref{sec:red-prec}) is performed in plaintext. We inject errors into ReLU during training by inverting outputs at a specified rate (e.g., outputting $x$ when $x < 0$ should output 0), enabling the network to develop robustness for reduced-precision encrypted inference.

\section{Evaluation Results}

\subsection{Homomorphic Operation Performance}
\begin{figure}[htbp]
\centering
\includegraphics[width=0.95\linewidth]{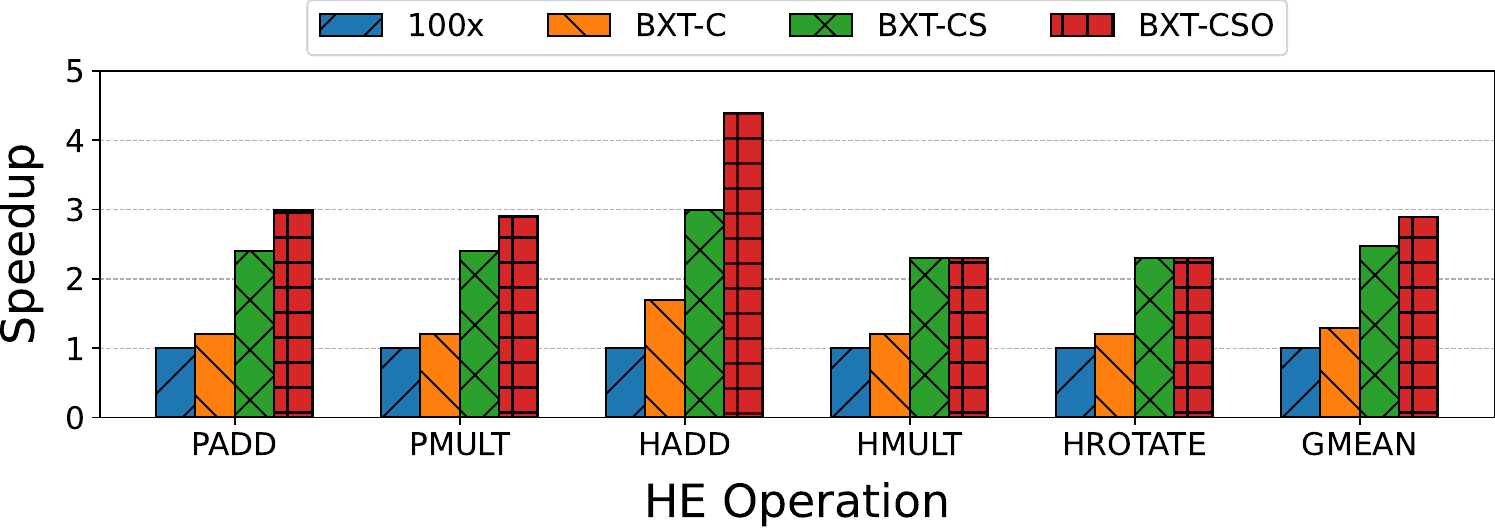}
\vspace{-0.5ex}
\caption{Execution time speedup over the 100x baseline~\cite{100x} for each homomorphic operation.}
\label{fig:Speedup_OPS}
\vspace{-3.5ex}
\end{figure}

We evaluate the speedup of each homomorphic operation in BXT relative to the 100x baseline~\cite{100x}, as shown in Figure~\ref{fig:Speedup_OPS}. BXT-C adds PRNG-based ciphertext compression that generates ciphertext matrices on-the-fly from a seed, BXT-CS additionally serializes ciphertexts into bit arrays, and BXT-CSO further applies ciphertext operation optimizations (delayed seed generation). We report results for PADD, PMULT, HADD, HMULT, and HROTATE.

Across all operations, BXT achieves consistent speedups over the baseline: 1.3$\times$ for BXT-C, 2.5$\times$ for BXT-CS, and 2.9$\times$ for BXT-CSO on average, with BXT-C alone ranging from 1.2$\times$ to 1.7$\times$. These gains mainly come from reducing matrix loads and stores via the PRNG engine, which is especially beneficial for memory-bound operations such as HADD, where PRNG avoids loading two matrices.

The layout modification in BXT-CS, which restructures ciphertext storage and access, nearly doubles the speedup over BXT-C by improving data layout and retrieval. Ciphertext operation optimizations in BXT-CSO further remove unnecessary loads and stores during the offline stage, but only for PADD, PMULT, and HADD, so HMULT and HROTATE do not benefit from BXT-CSO.

\begin{figure}[htbp]
\centering
\includegraphics[width=0.95\linewidth]{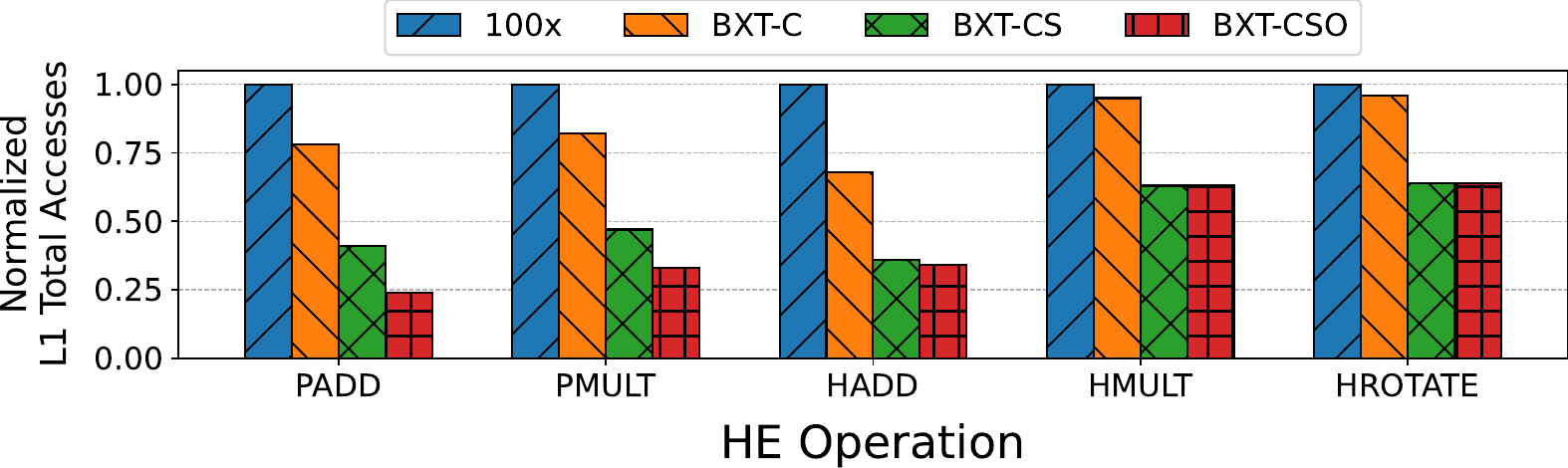} \\
\vspace{-1.5ex}
\includegraphics[width=0.95\linewidth]{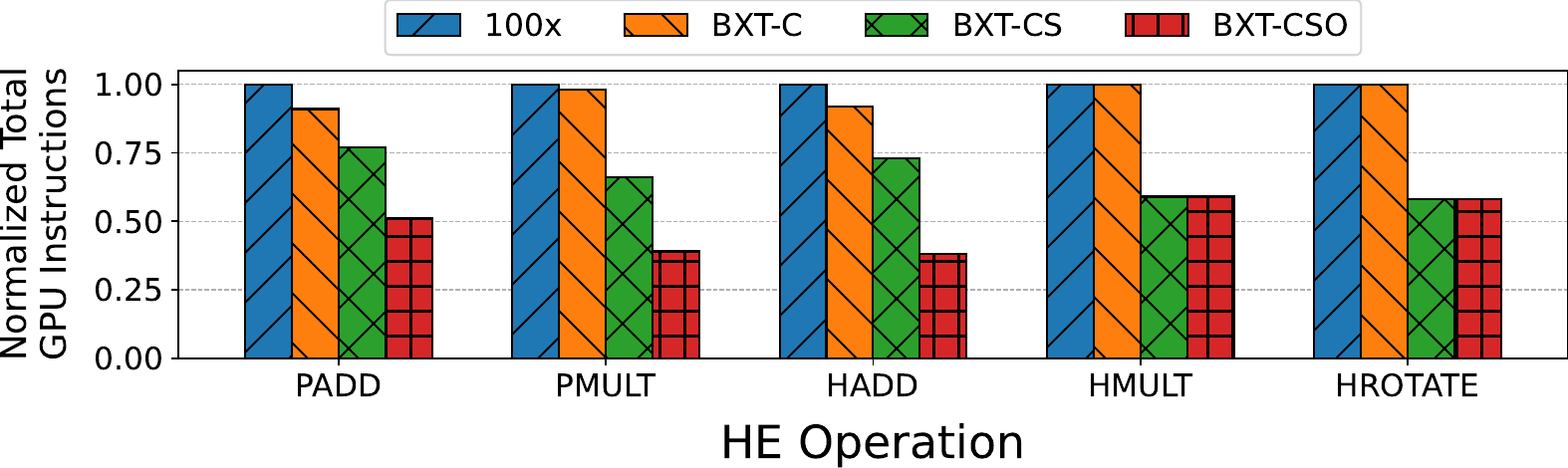}
\vspace{-0.5ex}
\caption{Total L1 accesses (top) and number of GPU instructions (bottom), normalized to the baseline 100x for each homomorphic operation.}
\label{fig:l1_total_accesses}
\vspace{-3ex}
\end{figure}

To understand the source of these gains, we also measured the number of instructions and cache accesses. Figure~\ref{fig:l1_total_accesses} (top) shows that each optimization reduces total L1 accesses, and Figure~\ref{fig:l1_total_accesses} (bottom) shows a similar decrease in GPU instructions; L2 accesses and misses follow the same trend, while L1/L2 miss rates remain roughly unchanged. Thus, the speedups primarily stem from fewer instructions and memory accesses without degrading cache locality. Furthermore, across all homomorphic operations with all BXT optimizations enabled (BXT-CSO), we achieve a geometric-mean arithmetic intensity increase of 1.5$\times$ over the baseline, indicating better utilization of memory bandwidth in these memory-bound uFHE workloads.

\subsection{Machine Learning Application Results}

\subsubsection{Performance Speedup}

\begin{figure}[htbp]
\vspace{-1ex}
\centering
\includegraphics[width=0.95\linewidth]{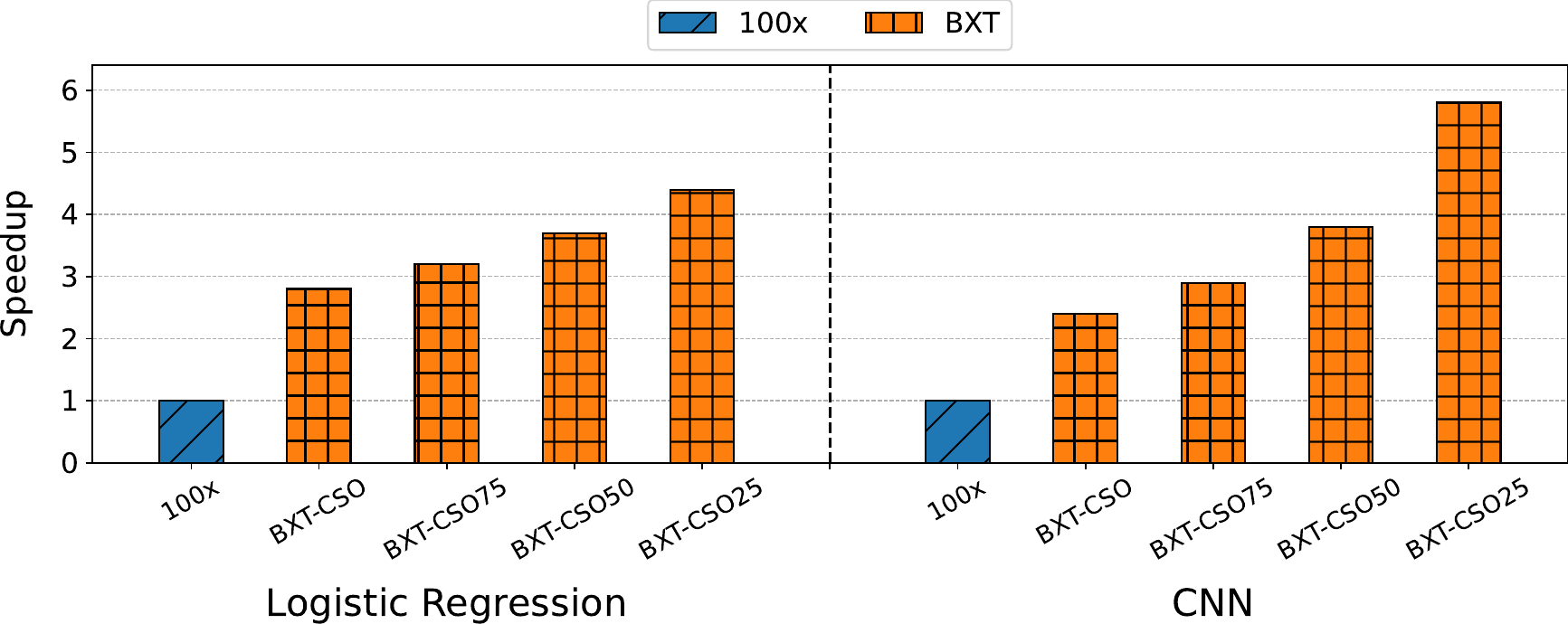} 
\vspace{-2ex}
\caption{Speedup for Logistic Regression (left) and CNN (right) with varying precision.} 
\label{fig:LR_CNN}
\vspace{-1ex}
\end{figure}

Figure~\ref{fig:LR_CNN} shows speedups for Logistic Regression and CNN. BXT-CSO achieves 2.8$\times$ (LR) and 2.4$\times$ (CNN) at full precision. With digit pruning, BXT-CSO50 (50\% precision) reaches 3.8$\times$ (LR/CNN), while BXT-CSO25 (25\% precision) achieves 4.4$\times$ (LR) and 5.8$\times$ (CNN).

\textbf{Arithmetic vs.\ ReLU Breakdown (CNN):} ReLU operations dominate CNN latency due to costly ciphertext comparisons that require many ciphertext multiplications. All BXT variants improve both arithmetic and ReLU performance, and with aggressive pruning (BXT-CSO50), ReLU speedup climbs to 4.2$\times$ while still accounting for most of the latency, confirming that optimizing ReLU is critical for uFHE performance.

\subsubsection{Precision Reduction Impact on Accuracy}

Table~\ref{tab:LR_CNN_Accuracy} presents the accuracy results for logistic regression and CNN workloads at various levels of precision reduction. For logistic regression, reducing precision to 75\% and 50\% has little impact on accuracy, while 25\% precision causes a drop of more than 3\%. Thus, maintaining at least 50\% precision is advantageous, providing up to 3.8$\times$ speedup without sacrificing accuracy.
For CNN, fault-aware training mitigates the impact of reduced precision, limiting the accuracy drop to less than 1\% at 50\% precision and less than 2\% at 25\% precision.

\begin{table}[htbp]
\vspace{-2ex}
\caption{Accuracy results for Logistic Regression and CNN models at varying levels of precision.}
\label{tab:LR_CNN_Accuracy}
\centering
  \begin{tabular}{|c|c|c|}
    \hline
     \makecell{\textbf{Comparison} \\ \textbf{Precision}} & \makecell{\textbf{Logistic Regression} \\ \textbf{Accuracy (\%)}} & \makecell{\textbf{CNN} \\ \textbf{Accuracy (\%)}} \\ 
     \hline \hline 
     100\% & 95.81 & 98.56 \\
     \hline
     75\% & 95.81 & 98.32 \\
     \hline
     50\% & 95.87 & 97.83 \\
     \hline
     25\% & 92.49 & 96.62 \\
     \hline
  \end{tabular}
  \vspace{-1ex}
\end{table}

\begin{figure}[htbp]
\vspace{-2ex}
\centering
\includegraphics[width=0.9\linewidth]{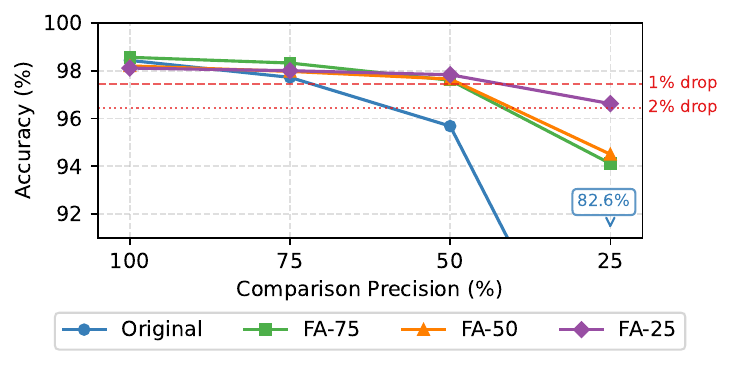}
\vspace{-2.5ex}
\caption{CNN accuracy with fault-aware training. FA-XX: trained at XX\% precision.}
\label{fig:cnn_fa_acc}
\vspace{-2ex}
\end{figure}

Figure~\ref{fig:cnn_fa_acc} shows the impact of different training strategies on CNN accuracy at reduced precision. Without fault-aware training, reducing precision to 25\% causes an accuracy drop of over 15\%. Conversely, the FA-25 model, trained with fault-aware training at 25\% precision, experiences a less significant accuracy drop of under 2\%. This reduction could be tolerable, considering that with 25\% comparison precision, the model achieves a speedup of more than 5$\times$. This demonstrates that fault-aware training is effective in enabling a larger degree of ciphertext digit reduction without sacrificing much accuracy.

\subsection{Comparison with Other Methods}
\label{sec:comparo}

\begin{figure}[htbp]
\centering
\includegraphics[width=0.95\linewidth]{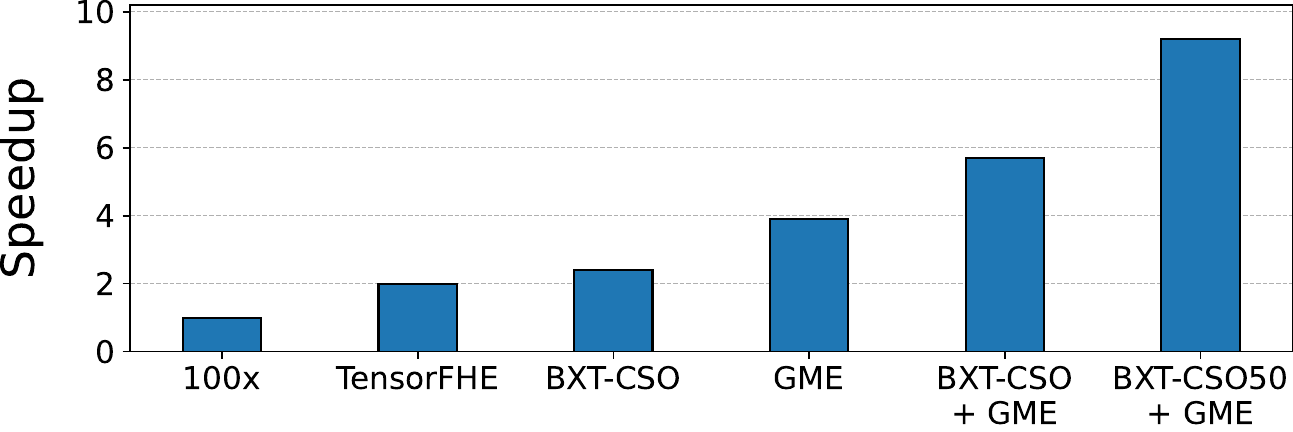}
\vspace{-2ex}
\caption{Comparison of the speedup achieved by TensorFHE, BXT-CSO, GME, BXT-CSO + GME, and BXT-CSO50 + GME relative to the baseline (100x) for CNN applications.}
\label{fig:speedup_CNN_comparison}
\vspace{-1ex}
\end{figure}

Figure~\ref{fig:speedup_CNN_comparison} shows the speedup of our schemes, TensorFHE~\cite{TensorFHE}, and GME~\cite{gme2023} relative to the 100x baseline~\cite{100x} for CNN applications. BXT-CSO, which uses all optimizations except comparison-precision reduction, outperforms TensorFHE but not GME. Because our optimizations are orthogonal to those of GME, combining BXT-CSO with GME yields a 5.7$\times$ speedup over the baseline, corresponding to a 1.5$\times$ speedup over GME.

With a 50\% comparison-precision reduction, BXT-CSO50 combined with GME achieves a 9.2$\times$ speedup with less than 1\% accuracy loss, corresponding to a 2.3$\times$ speedup over GME. This result shows that BXT integrates easily with complementary GPU acceleration techniques while maintaining high efficiency and minimal accuracy impact.

\section{Discussion}

\noindent\textbf{Generalizability.} Compression and serialization apply broadly to FHE schemes but are most effective for memory-bound uFHE. Delayed seed generation is workload-dependent, determined by data formats and operation patterns. Ciphertext pruning is particularly effective for ReLU and other activation functions where LSB accuracy is less critical, and it also extends to non-linear operations like division and sorting.

\noindent\textbf{Microarchitecture Integration \& Overhead.} BXT's serialization requires a bit-to-integer converter between L2 and L1 cache (2-cycle latency assumed). Sensitivity analysis (2–-32 cycles) shows negligible performance impact due to the memory-bound nature of uFHE. PRNG hardware (40-cycle latency per SM) is available in modern GPUs~\cite{yuan2021analyzing}. The additional hardware support is limited in scope: the converter uses simple bit-packing logic, and PRNG units are already present in GPUs. Our techniques integrate into existing GPU architectures without invasive changes, unlike custom NoCs (GME~\cite{gme2023}) or specialized functional units (CraterLake~\cite{CraterLake}).

\noindent\textbf{Compression vs.\ Generic Methods.} Generic compression (e.g., gzip, LZ4) fails on ciphertexts due to high entropy from encryption's diffusion property. BXT exploits FHE-specific structure: (1) seeded PRNG regeneration leverages BGV's deterministic polynomial generation, and (2) serialization removes unused bits based on known modulus sizes. Domain-specific knowledge is essential for effective ciphertext compression.
\section{Conclusion}

In this study, we addressed the substantial computational demands of FHE by introducing a novel multi-level strategy for reducing memory usage in FHE computation, derived from algorithmic insights, application-specific considerations, and data access patterns. Through evaluation on CNN workloads, we achieved a notable speedup of 3.8$\times$ compared to the 100x baseline with less than 1\% accuracy reduction. This highlights the effectiveness of reducing memory accesses in highly memory-constrained workloads, leading to a significant acceleration in execution time.
\textcolor{black}{\section*{Acknowledgment}}

\textcolor{black}{We thank the reviewers for their valuable comments. This work was supported in part by the National Science Foundation under Grant No. 2413232. The views expressed are those of the authors and do not necessarily reflect those of the NSF.}


\begin{thebibliography}{10}

\bibitem{Gentry2012FHEPolylog}
C.~Gentry, S.~Halevi, and N.~Smart, ``Fully homomorphic encryption with polylog overhead,'' in {\em Advances in Cryptology - EUROCRYPT 2012}, vol.~7237, pp.~465--482, 2012.

\bibitem{AmazonScience}
{Amazon Science}, ``Machine learning models that act on encrypted data.'' \url{https://www.amazon.science/blog/machine-learning-models-that-act-on-encrypted-data}, Nov 2020.
\newblock Accessed: 2025-09-07.

\bibitem{TensorFHE}
S.~Fan, Z.~Wang, W.~Xu, R.~Hou, D.~Meng, and M.~Zhang, ``Tensorfhe: Achieving practical computation on encrypted data using gpgpu,'' in {\em Proc. 29th HPCA}, pp.~922--934, 2023.

\bibitem{reagen2021cheetah}
B.~Reagen, W.-S. Choi, Y.~Ko, V.~T. Lee, H.-H.~S. Lee, G.-Y. Wei, and D.~Brooks, ``Cheetah: Optimizing and accelerating homomorphic encryption for private inference,'' in {\em Proc. 27th HPCA}, pp.~26--39, IEEE, 2021.

\bibitem{brakerski2014leveled}
Z.~Brakerski, C.~Gentry, and V.~Vaikuntanathan, ``(leveled) fully homomorphic encryption without bootstrapping,'' {\em ACM Transactions on Computation Theory}, vol.~6, no.~3, pp.~1--36, 2014.

\bibitem{CKKS_scheme}
J.~H. Cheon, A.~Kim, M.~Kim, and Y.~Song, ``Homomorphic encryption for arithmetic of approximate numbers,'' in {\em Advances in Cryptology -- ASIACRYPT 2017}, pp.~409--437, 2017.

\bibitem{TFHE_scheme}
I.~Chillotti, N.~Gama, M.~Georgieva, and M.~Izabach{\`e}ne, ``Faster fully homomorphic encryption: Bootstrapping in less than 0.1 seconds,'' in {\em Advances in Cryptology -- ASIACRYPT 2016}, pp.~3--33, 2016.

\bibitem{iliashenko2021faster}
I.~Iliashenko and V.~Zucca, ``Faster homomorphic comparison operations for bgv and bfv,'' {\em Proc. on Privacy Enhancing Technologies}, vol.~2021, no.~3, pp.~246--264, 2021.

\bibitem{de2021does}
L.~de~Castro, R.~Agrawal, R.~Yazicigil, A.~Chandrakasan, V.~Vaikuntanathan, C.~Juvekar, and A.~Joshi, ``Does fully homomorphic encryption need compute acceleration?,'' {\em arXiv preprint arXiv:2112.06396}, 2021.

\bibitem{100x}
W.~Jung, S.~Kim, J.~H. Ahn, J.~H. Cheon, and Y.~Lee, ``Over 100x faster bootstrapping in fully homomorphic encryption through memory-centric optimization with gpus,'' {\em IACR TCHES}, vol.~2021, no.~4, p.~114–148, 2021.

\bibitem{mad23}
R.~Agrawal, L.~De~Castro, C.~Juvekar, A.~Chandrakasan, V.~Vaikuntanathan, and A.~Joshi, ``Mad: Memory-aware design techniques for accelerating fully homomorphic encryption,'' in {\em Proc. 56th MICRO}, p.~685–697, 2023.

\bibitem{accel-sim}
M.~Khairy, Z.~Shen, T.~M. Aamodt, and T.~G. Rogers, ``Accel-sim: An extensible simulation framework for validated gpu modeling,'' in {\em Proc. 47th ISCA}, pp.~473--486, 2020.

\bibitem{gme2023}
K.~Shivdikar, Y.~Bao, R.~Agrawal, M.~Shen, G.~Jonatan, E.~Mora, A.~Ingare, N.~Livesay, J.~L. Abell\'{A}N, J.~Kim, A.~Joshi, and D.~Kaeli, ``Gme: Gpu-based microarchitectural extensions to accelerate homomorphic encryption,'' in {\em Proc. 56th MICRO}, p.~670–684, 2023.

\bibitem{HE-Booster}
Z.~Wang, P.~Li, R.~Hou, Z.~Li, J.~Cao, X.~Wang, and D.~Meng, ``He-booster: An efficient polynomial arithmetic acceleration on gpus for fully homomorphic encryption,'' {\em IEEE Trans. on Parallel and Distributed Systems}, vol.~34, no.~4, pp.~1067--1081, 2023.

\bibitem{kim2024cheddarswiftfullyhomomorphic}
W.~Choi, J.~Kim, and J.~H. Ahn, ``Cheddar: A swift fully homomorphic encryption library designed for gpu architectures,'' in {\em Proc. 31st ASPLOS, Volume 1}, p.~35–49, 2025.

\bibitem{Deng2024Trinity}
X.~Deng, S.~Fan, Z.~Hu, Z.~Tian, Z.~Yang, J.~Yu, D.~Cao, D.~Meng, R.~Hou, M.~Li, Q.~Lou, and M.~Zhang, ``Trinity: A general purpose fhe accelerator,'' in {\em Proc. 57th MICRO}, pp.~338--351, 2024.

\bibitem{wang2024chameleonefficientfhescheme}
Z.~Wang, H.~He, L.~Zhao, P.~Li, Z.~Li, D.~Meng, and R.~Hou, ``Chameleon: An efficient fhe scheme switching acceleration on gpus,'' {\em IEEE Trans. on Parallel and Distributed Systems}, vol.~36, no.~11, pp.~2264--2280, 2025.

\bibitem{Ozgun2022EfficientNTT}
O.~\"{O}zerk, C.~Elgezen, A.~C. Mert, E.~\"{O}zt\"{u}rk, and E.~Sava\c{s}, ``Efficient number theoretic transform implementation on gpu for homomorphic encryption,'' {\em J. Supercomput.}, vol.~78, p.~2840–2872, Feb. 2022.

\bibitem{SHARP_ISCA23}
J.~Kim, S.~Kim, J.~Choi, J.~Park, D.~Kim, and J.~H. Ahn, ``Sharp: A short-word hierarchical accelerator for robust and practical fully homomorphic encryption,'' in {\em Proc. 50th ISCA}, 2023.

\bibitem{ARK}
J.~Kim, G.~Lee, S.~Kim, G.~Sohn, M.~Rhu, J.~Kim, and J.~H. Ahn, ``Ark: Fully homomorphic encryption accelerator with runtime data generation and inter-operation key reuse,'' in {\em Proc. 55th MICRO}, pp.~1237--1254, 2022.

\bibitem{BTS}
S.~Kim, J.~Kim, M.~J. Kim, W.~Jung, J.~Kim, M.~Rhu, and J.~H. Ahn, ``Bts: An accelerator for bootstrappable fully homomorphic encryption,'' in {\em Proc. 49th ISCA}, p.~711–725, 2022.

\bibitem{F1}
N.~Samardzic, A.~Feldmann, A.~Krastev, S.~Devadas, R.~Dreslinski, C.~Peikert, and D.~Sanchez, ``F1: A fast and programmable accelerator for fully homomorphic encryption,'' in {\em Proc. 54th MICRO}, p.~238–252, 2021.

\bibitem{CraterLake}
N.~Samardzic, A.~Feldmann, A.~Krastev, N.~Manohar, N.~Genise, S.~Devadas, K.~Eldefrawy, C.~Peikert, and D.~Sanchez, ``Craterlake: A hardware accelerator for efficient unbounded computation on encrypted data,'' in {\em Proc. 49th ISCA}, p.~173–187, 2022.

\bibitem{FxHENN}
Y.~Zhu, X.~Wang, L.~Ju, and S.~Guo, ``Fxhenn: Fpga-based acceleration framework for homomorphic encrypted cnn inference,'' in {\em Proc. 29th HPCA}, pp.~896--907, 2023.

\bibitem{rasoul2024compressingfhe}
R.~A. Mahdavi, A.~Diaa, and F.~Kerschbaum, ``He is all you need: Compressing fhe ciphertexts using additive he,'' {\em arXiv preprint arXiv:2303.09043}, 2023.

\bibitem{yudha2024boostcom}
A.~W.~B. Yudha, J.~Xue, Q.~Lou, H.~Zhou, and Y.~Solihin, ``Boostcom: Towards efficient universal fully homomorphic encryption by boosting the word-wise comparisons,'' in {\em Proc. PACT 2024}, p.~121–132, 2024.

\bibitem{Chimera}
C.~Boura, N.~Gama, M.~Georgieva, and D.~Jetchev, ``Chimera: Combining ring-lwe-based fully homomorphic encryption schemes,'' {\em Journal of Mathematical Cryptology}, vol.~14, no.~1, pp.~316--338, 2020.

\bibitem{lu2021pegasus}
W.-j. Lu, Z.~Huang, C.~Hong, Y.~Ma, and H.~Qu, ``Pegasus: bridging polynomial and non-polynomial evaluations in homomorphic encryption,'' in {\em Proc. 42nd S\&P}, pp.~1057--1073, IEEE, 2021.

\bibitem{Cooley1965FFT}
J.~W. Cooley and J.~W. Tukey, ``An algorithm for the machine calculation of complex fourier series,'' {\em Mathematics of Computation}, vol.~19, no.~90, pp.~297--301, 1965.

\bibitem{HElib}
S.~Halevi and V.~Shoup, ``Design and implementation of helib: a homomorphic encryption library.'' Cryptology ePrint Archive, Paper 2020/1481, 2020.

\bibitem{OpenFHE}
A.~Al~Badawi, J.~Bates, F.~Bergamaschi, D.~B. Cousins, S.~Erabelli, N.~Genise, S.~Halevi, H.~Hunt, A.~Kim, Y.~Lee, Z.~Liu, D.~Micciancio, I.~Quah, Y.~Polyakov, S.~R.V., K.~Rohloff, J.~Saylor, D.~Suponitsky, M.~Triplett, V.~Vaikuntanathan, and V.~Zucca, ``Openfhe: Open-source fully homomorphic encryption library,'' in {\em Proc. 10th WAHC}, p.~53–63, 2022.

\bibitem{Markidis2018TensorCore}
S.~Markidis, S.~W.~D. Chien, E.~Laure, I.~B. Peng, and J.~S. Vetter, ``{NVIDIA Tensor Core Programmability, Performance \& Precision},'' in {\em Proc. IPDPSW 2018}, pp.~522--531, May 2018.

\bibitem{michaud2023precisionML}
E.~J. Michaud, Z.~Liu, and M.~Tegmark, ``Precision machine learning,'' {\em Entropy}, vol.~25, no.~1, 2023.

\bibitem{zahid2020fat}
U.~Zahid, G.~Gambardella, N.~J. Fraser, M.~Blott, and K.~Vissers, ``Fat: Training neural networks for reliable inference under hardware faults,'' in {\em 2020 IEEE International Test Conference}, pp.~1--10, IEEE, 2020.

\bibitem{yuan2021analyzing}
S.~Yuan, A.~W. Baskara~Yudha, Y.~Solihin, and H.~Zhou, ``Analyzing secure memory architecture for gpus,'' in {\em Proc. ISPASS 2021}, pp.~59--69, 2021.

\bibitem{lecun1998mnist}
Y.~LeCun, ``The mnist database of handwritten digits,'' {\em http://yann.lecun.com/exdb/mnist/}, 1998.

\bibitem{cryptodl}
E.~Hesamifard, H.~Takabi, and M.~Ghasemi, ``Cryptodl: Deep neural networks over encrypted data,'' {\em arXiv preprint arXiv:1711.05189}, 2017.

\end{thebibliography}

\end{document}